# An Efficient Algorithm for Mining Multilevel Association Rule Based on Pincer Search

Dr. Pratima Gautam[1], Rahul Shukla[2]

[1] Computer Application, MANIT
Bhopal, M.P. 462032, India
Mechanical Department, BIST
Bhopal, M.P., 462021, India

**Abstract**
Discovering frequent itemset is a key difficulty in significant data mining applications, such as the discovery of association rules, strong rules, episodes, and minimal keys. The problem of developing models and algorithms for multilevel association mining poses for new challenges for mathematics and computer science. In this paper, we present a model of mining multilevel association rules which satisfies the different minimum support at each level, we have employed princer search concepts, multi-level taxonomy and different minimum supports to find multilevel association rules in a given transaction data set. This search is used only for maintaining and updating a new data structure. It is used to prune early candidates that would normally encounter in the top-down search. A main characteristic of the algorithms is that it does not require explicit examination of every frequent itemsets, an example is also given to demonstrate and support that the proposed mining algorithm can derive the multiple-level association rules under different supports in a simple and effective manner

## 1. Introduction

The vision of exploring and discovering new and hidden patterns in data has been given a variety of names counting data mining, knowledge extraction, information discovery, information harvesting, data archaeology, and data pattern processing. The term data mining has been generally used by statisticians, data analysts, and association information systems communities [11], [16]. The goal of data mining is to discover important associations among items such that the presence of some items in a transaction will imply the presence of some other items. To accomplish this purpose, Agrawal and his co-workers proposed several mining algorithms based on the concept of large itemsets to find association rules in transaction data [2], [3]. They separated the mining process into two phases, in the first phase, frequent (large) itemsets are found based on the counts by scanning the transaction data. In the second phase, association rules were induced from the large itemsets found in the first phase. Charge of association rule mining is to find hidden, earlier than unknown and potentially helpful information in huge amount of data [13], [15]. The difficulty of discovering frequent patterns has established an enormous agreement of mind [12]. The problem is broadly known as market basket analysis; for example a supermarket database at someplace for the set of items purchased by a customer on a single visit to a store is recorded as a transaction [1], [4]. The superstore managers might be involved in ruling associations among the items purchased mutually in one transaction. The result of association analysis is burly association rules, which are rules satisfying a minimal support and a minimal confidence threshold. The minimal support and the minimal confidence are input parameters for association analysis [15].Basically association rules follow apriori principle, but apriori algorithm operates in a bottom-up, breadth-first search method. The computation start from the smallest set of frequent itemsets and moves upward till it reaches the largest frequent itemset. This method increases number of database scans and algorithm has to go much iteration; as a result, the performance decreases [16]. To overcome this problem bi-directional search is developed, which takes both bottom-up as well as the top down process. The princer search algorithm based on this principle. We apply princer search method for multilevel association rule. Currently multi-level datasets are more common in several domains [14]. Through this increase in usage, there is a big demand for techniques to discover multi-level and cross-level association rules and also techniques to measure interestingness of rules derived from multi-level datasets [ 17][ 19].The multilevel rules are based on analyses of the rules mined from atomic level, in its place of using traditional methods which extract from database again. Consequently, it has an excellent probability to reduce the computational difficulty, number of scans and I/O cost. In totality, our proposed method supports hierarchies which can be dynamically constructed according to the input of user knowledge. The encoding scheme in [18] is adapted to collection the association rules generated from atomic level. Here we propose the Princer search and Boolean method, which solves the problem of discovering large itemsets [20], [19]. It uses the new procedure for candidate generation, reduce the number of scan, I/O cost and reduce CPU overhead which is more capable than the suitable process from the obtainable algorithm.

## 2. Multilevel Association Rule Mining

We can mine multilevel association rules professionally using concept hierarchies, which defines a series of mappings from a set of low level concepts to higher-level, more general concepts. On lowest levels, it strength will be that no rules may match the constraints [9]. At highest levels, rules can be tremendously universal. Usually, a top-down proceed is used where the support threshold varies from level to level (support is abridged leaving from higher to lower levels) [10]. In a lot of the applications of single-level association rule mining, the items enclosed in an itemsets could potentially be hierarchically prearranged where primitive level concepts can be universal to higher levels while the more specific items to lower levels of the hierarchy [7]. Through forming such a concept hierarchy, a course of discovering association rules at multiple concept levels progressively deepens the knowledge mining procedure for finding more important and advanced knowledge from the data [17],[18]. Toward demonstrate, here is an example of a concept hierarchy.

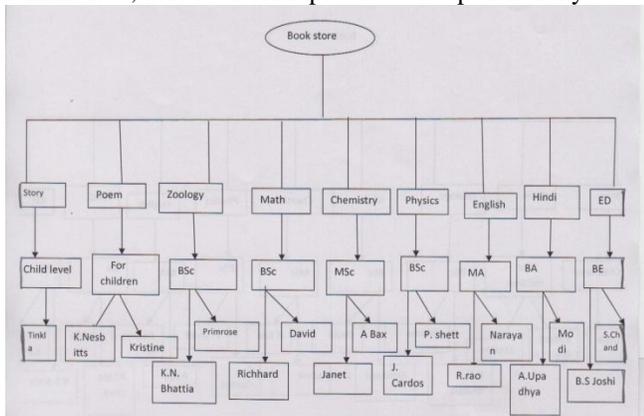

Fig. 1 Concept Hierarchy Tree Example

Each node in the concept-hierarchy tree in Figure 1 represents one item in the itemset (except for the root node "Book Store" since it represents all items). intended for, the item `story child level thinkle' is encoded as `A11' in which the first digit, `A', represents `story' at level-1, the second, `1', for ` child level' at level-2, and the third, `1', for the 'writer` at level-3. Similar to [2] [3], constant items (i.e., items with the same encoding) at any level will be treated as one item in one transaction.

## 3. The Basic Pincer-Search Algorithm

$L_0 :=$ null; $k := 1$; $C_1 := \{\{ i \} \mid i$ belong to $I_0\}$
MFCS: $= I_0$; MFS: $=$ null;
while $C_k$ != null
  read database and count supports for $C_k$ and MFCS
  remove frequent itemsets from MFCS and add them to MFS
   determine frequent set $L_k$ and infrequent set $S_k$
  use $S_k$ to update MFCS
 generate new candidate set $C_{k+1}$ (join, recover, and prune)
  $k := k +1$
 return MFS

## 4. Proposed Method

In this part, we explain the main skin of our account which overcomes the problems of I/O cost, CPU overhead, database scan as discovering frequent itemsets in large database [12]. We use Han and Fu's encoding scheme, as described in [1] to represent nodes in predefined taxonomies for mining multilevel rules. Nodes are encoded with respect to their positions in the hierarchy using sequences of numbers and the symbol "*". It uses Boolean form which is show presence and absence of itemsets and also uses top-down progressive deepening method [13], [17]. Because analysis of earlier algorithms we are using the idea of our algorithm that was given [16], [18] and I am applying this method into multilevel association rule mining.

The algorithm consists of following steps:
Step-1:
Encode taxonomy using a sequence of numbers and the symbol ''*'', with the $l$th number representing the branch number of a certain item at levels.
Step-2: Set k= 1, where l is used to store the level number being processed whereas l∈ {1, 2, 3} (As we consider up to 3-levels of hierarchies).
Step-3:
Transforming the transaction databases into the boolean form [13]. Here 0 represent the absence of itemsets and 1 represent presence of itemsets.
Step-4:
Set user defines minimum support on current level.
Step-5:
Count the itemsets according the occurrences of itemsets in the transaction dataset. After that evaluate predefine minimum support threshold.
Step-6:
Determine frequent itemset L and infrequent itemset S.
Step-7:
Use $S$ to update MFCS // Maximal frequent candidate set
Step-8
Generate new candidate set $C_{k+1}$ (join, recover, and prune)

Step-9
Generate k+1; (Increment l value by 1; i.e., l = 2, 3) itemset from K and go to step-4 (for repeating the intact processing for next level).

## 4.1 An illustrative example

An illustrative example is given to understand well the concept of the proposed method and algorithm and how the process of the generating multilevel association rule mining is performed step by step [17]. The process is started from a given transactional database as shown in Table 1.

Table-1 Transaction Database

| Tid | Itemes |
|---|---|
| T1 | A11, E11, F11, H11 |
| T2 | B11, D11, H12 |
| T3 | D12, E11, G11 |
| T4 | C11 |
| T5 | E11, F12, G12 |
| T6 | B11, C11, D12 |
| T7 | B12, F11, G12, I11 |
| T8 | E12, H11 |
| T9 | C12, D12, E12, G12 |
| T10 | C11, D12, E11, G11 |
| T11 | C11, D11, E12, G12 |
| T12 | E11, F11, H11 |
| T13 | B12,C12, E12,F12 |
| T14 | A11, C12, D11,E11,G11 |
| T15 | B11, C11,I11 |

Table-1[b] Encoded Transaction Database

| Codes | Items |
|---|---|
| A** | Story book |
| B** | Poem book |
| C** | Zoology book |
| D** | Math book |
| E** | Chemistry book |
| F** | Physics book |
| G** | English book |
| H** | Hindi book |
| I** | Dawning book |
| A1* | Child Story book |
| B1* | Child Poem book |
| C1* | Bsc Exam Zoology book |
| D1* | Bsc Exam Math book |
| E1* | Bsc Exam Chemistry book |
| F1* | Bsc Exam Physics book |
| G1* | Elective Sub. English for Bsc |
| H1* | Elective Sub. Hindi for Bsc |
| I1* | BE Drawing book |
| A11 | Child Story book Tinkla |
| B11 | Child Poem book Kenn |
| B12 | Child Poem book Kristin |
| C11 | Bsc Exam Zoology book Bhatia |
| C12 | Bsc Exam Zoology book Primrose |
| D11 | Bsc Exam Math book Richard |
| D12 | Bsc Exam Math book David |
| E11 | Bsc Exam Chemistry book Janet |
| E12 | Bsc Exam Chemistry book Ad Bax |
| F11 | Bsc Exam Physics book Cardoso |
| F12 | Bsc Exam Physics book P. Sheth |
| G11 | Elective Sub. English for Bsc Narayan |
| G12 | Elective Sub. English for Bsc Rja Rao |
| H11 | Elective Sub. Hindi for Bsc Upadhya |
| H12 | Elective Sub. Hindi for Bsc P.K. Modi |
| I11 | S. Chand |

Table-3 Transform the database into Boolean form.

| id | A** | B** | C** | D** | E** | F** | G** | H** | I** |
|---|---|---|---|---|---|---|---|---|---|
| T1 | 1 | 0 | 0 | 0 | 1 | 1 | 0 | 1 | 0 |
| T2 | 0 | 1 | 0 | 1 | 0 | 0 | 0 | 1 | 0 |
| T3 | 0 | 0 | 0 | 1 | 1 | 0 | 1 | 0 | 0 |
| T4 | 0 | 0 | 1 | 0 | 0 | 0 | 0 | 0 | 0 |
| T5 | 0 | 0 | 0 | 0 | 1 | 1 | 1 | 0 | 0 |
| T6 | 0 | 1 | 1 | 1 | 0 | 0 | 0 | 0 | 0 |
| T7 | 0 | 1 | 0 | 0 | 0 | 1 | 1 | 0 | 1 |
| T8 | 0 | 0 | 0 | 0 | 1 | 0 | 0 | 1 | 0 |
| T9 | 0 | 0 | 1 | 1 | 1 | 0 | 1 | 0 | 0 |
| T10 | 0 | 0 | 1 | 1 | 1 | 0 | 1 | 0 | 0 |
| T11 | 0 | 0 | 1 | 1 | 1 | 0 | 1 | 0 | 0 |
| T12 | 0 | 0 | 0 | 0 | 1 | 1 | 0 | 1 | 0 |
| T13 | 0 | 1 | 1 | 0 | 1 | 1 | 0 | 0 | 0 |
| T14 | 1 | 0 | 1 | 1 | 1 | 0 | 1 | 0 | 0 |
| T15 | 0 | 1 | 1 | 0 | 0 | 0 | 0 | 0 | 1 |

**Level-1**
**Min_Support = 3**
C1= {A**, B**, C**, D**, E**, F**, G**, H**, I**}

MFCS: - {A**, B**, C**, D**, E**, F**, G**, H**, I**}
MFS = φ
**Pass One: Database is read to count the support as follows:**

{A**} = 2, {B**} = 5, {C**} = 8, {D**} = 7, {E**} = 10, {F**} = 5, {G**} = 7, {H**} =4, {I**} = 2
{A**, B**, C**, D**, E**, F**, G**, H**, I**}→ 0
So MFCS: = {A**, B**, C**, D**, E**, F**, G**, H**, I**} and MFS= φ;

$L^1_1$= {{B**}, {C**}, {D**}, {E**}, {F**}, {G**}, {H**}} //$L^1_1$ represent frequent 1-itemsets at level1

S: = {A**}, {I**} // S represents infrequent itemsets

At this stage we call the MFCS-gen to update MFCS. For {A**} in S and for {B**, C**, D**, E**, F**, G**, H**, I**} in MFCS, we get the new element in MFCS as {B**, C**, D**, E**, F**, G**, H**, I**}

For {I**} in S and for {B**, C**, D**, E**, F**, G**, H**, I**} in MFCS, we get the new elements in MFCS as {B**, C**, D**, E**, F**, G**, H**}

After that we compare the support count threshold into itemsets frequency and find frequent 1-itemset.

**Frequent_1_itemsets**

$C^1_1$= {B**}, {C**}, {D**}, {E**}, {F**}, {G**}, {H**}
We generate the 2-candidate itemsets

**Frequent_2_itemsets**

$C^2_1$= {B**, C**}, {B**, D**}, {B**, E**}, {B**, F**}, {B**, G**}, {B**, H**}, {C**, D**}, {C**, E**}, {C**, F**}, {C**, G**}, {C**, H**}, {D**, E**}, {D**, F**}, {D**, G**}, {D**, H**}, {E**, F**}, {E**, G**}, {E**, H**}, {F**, G**}, {F**, H**}, {G**, H**}

**Pass Two: Read the database to count the support of elements in $C^2_1$ and MFCS as given below:**

{B**, C**}= 3, {B**, D**}= 2, {B**, E**}= 1,{B**, F**}=2,{B**, G**}=1, {B**, H**}=1, {C**, D**}=5 {C**, E**}=5, {C**, F**}= 1,{C**, G**}= 4,{C**, H**}= 0,{D**, E**}= 5,{D**, F**}= 1,{D**, G**}= 5, {D**, H**}=1, {E**, F**}=4, {E**, G**}=6, {E**, H**}=3, {F**, G**}=2, {F**, H**}=2, {G**, H**}=0
So MFCS: = {B**}, {C**}, {D**}, {E**}, {F**}, {G**}, {H**} and MFS= φ;

$L^2_1$: = {B**, C**}, {C**, D**}, {C**, E**}, {C**, G**}, {D**, E**}, {D**, G**}, {E**, F**}, {E**, G**}, {E**, H**} // $L^2_1$ represent frequent2-itemsets at level1

S: = {B**, D**} {B**, E**}, {B**, F**}, {B**, G**}, {B**, H**}, {C**, F**}, {C**, H**},{D**, F**},{D**, H**},{F**, G**}, {F**, H**}, {G**, H**} //S represents infrequent itemsets
At this stage we call the MFCS-gen to update MFCS for {B**, D**} {B**, E**}, {B**, F**}, {B**, G**}, {B**, H**}, {C**, F**}, {C**, H**},{D**, F**},{D**, H**},{F**, G**}, {F**, H**}, {G**, H**}and get new elements in MFCS.

**After that we generate 3-itemsets**

$C^3_1$= {B**, C**, D**}, {B**, C**, E**}, {B**, C**, F**}, {B**, C**, G**}, {B**, C**, H**}, {C**, D**, E**},{C**, D**, F**},{C**, D**, G**},{C**, D**, H**},{D**, E**, G**},{D**, E**, H**},{E**, F**,G**},{E**, F**,H**}

**Read the database to count the support of elements in $C^3_1$ and MFCS as given below:**

{B**, C**, D**}=1, {B**, C**, E**}=0, {B**, C**, F**}=1, {B**, C**, G**}=0, {B**, C**, H**}=0, {C**, D**, E**}=4,{C**, D**, F**}=0,{C**, D**, G**}=4, {C**, D**, H**}=0,{D**, E**, G**}=5,{D**, E**, H**}=0,{E**, F**,G**}=1,{E**, F**,H**}=1

So MFCS: = {B**}, {C**}, {D**}, {E**}, {G**} and MFS= φ;

$L^3_1$:= {C**, D**, E**}, {C**, D**, G**}, {D**, E**, G**} // $L^3_1$ represent frequent 3-itemsets at level1

S: = {B**, C**, D**}, {B**, C**, E**}, {B**, C**, F**}, {B**, C**, G**}, {B**, C**, H**}, {C**, D**, F**}{C**, D**, H**},{D**, E**, H**},{E**, F**,G**},{E**, F**,H**}

At this stage we call the MFCS-gen to update MFCS for {B**, C**, D**}, {B**, C**, E**}, {B**, C**, F**}, {B**, C**, G**}, {B**, C**, H**}, {C**, D**, F**}{C**, D**, H**},{D**, E**, H**},{E**, F**,G**},{E**, F**,H**} and get new elements in MFCS.

We also find 4-itemset at level-1

$L^4_1$={C**, D**, E**, G**} =4
We take the C**, D**, E**, G** itemsets and go the next level.

**Level -2
Min_Support = 2.0
1-itemsets**

$L^1_2$:= C1*=8, D1=7, E1*=10, F1*= 5, G1*=7
So MFCS: = {C1*}, {D1*}, {E1*}, {F1*}, {G1*} and MFS= ϕ;
All items making 2-itemsets

**2-itemsets**

{C1*, D1*} = 5, {C1*, E1*} = 5, {C1*, F1*} =1, {C1*, G1*} = 4, {D1*, E1*} =5, {D1*, F1*} =1, {D1*, G1*} = 5, {E1*, F1*} = 4, {E1*, G1*} = 5, {F1*, G1*} = 2

$L^2_2$:= {C1*, D1*}, {C1*, E1*}, {C1*, G1*}, {D1*, E1*}, {D1*, G1*}, {E1*, F1*}, {E1*, G1*}, {F1*, G1*}
// $L^2_2$ represent frequent 2-itemsets at level-2

S: = {C1*, F1*}, {D1*, F1*} // Infrequent itemset
At this stage we call the MFCS-gen to update MFCS for {C1*, F1*}, {D1*, F1*} and get new elements in MFCS.

And generate 3-itemsets.

**3-itemsets**

$L^3_2$:= {C1*, D1*, E1*} = 4, {C1*, D1*, G1*} = 4, {D1*, E1*, G1*} = 5

Again all itemsets making 4-itemsets

**4-itemsets**

$L^4_2$:= {C1*, D1*, E1*, G1*} = 4
C1*, D1*, E1*, G1* itemsets carry for next level

**Level-3
Min_Support = 2
1-itemsets**

C11= 5, C12= 3, D11=3, D12=4, E11=6, E12=10, G11=3, G12=4
So MFCS: = {C11}, {C12}, {D11}, {D12}, {E11}, {E12}, {G11}, {G12} and MFS= ϕ;

$L^1_3$:= {C11}, {C12}, {D11}, {D12}, {E11}, {E12}, {G11}, {G12}// frequent itemsets

All itemsets making 2-itemsets

**2-itemsets**

{C11, C12}=0, {C11, D11}=1,{C11, D12}= 2,{C11, E11}=1, {C11, E12}=1,{C11, G11}=1,{C11, G12}=,1{D11, D12}= 0, {D11, E11} = 1, {D11, E12} = 1, {D11, G11} = 1, {D11, G12} = 1, {D12, E11} = 2,{D12, E12}= 1, {D12, G11} = 2, {D12, G12} = 1, {E11, E12} = 0, {E11, G12} =2, {E12, G11} =0, {E12, G12} =2, {G11, G12} =0.

$L^2_3$:= {C11, D12}, {D12, E11}, {D12, G11}, {E11, G12}, {E12, G12}// Frequent itemsets at level-3

S:= {C11, C12}, {C11, D11},{C11, E11}, {C11, E12},{C11, G11},{C11, G12},{D11, D12}, {D11, E11}, {D11, E12}, {D11, G11}, {D11, G12} , {D12, E12}, {D12, G12} , {E11, E12} , {E12, G11}, , {G11, G12}.

At this stage we call the MFCS-gen to update MFCS, {C11, C12}, {C11, D11},{C11, E11}, {C11, E12},{C11, G11},{C11, G12},{D11, D12}, {D11, E11}, {D11, E12}, {D11, G11}, {D11, G12} , {D12, E12}, {D12, G12} , {E11, E12} , {E12, G11}, , {G11, G12}.and get new elements in MFCS.
After that {C11, D12}, {D12, E11}, {D12, G11}, {E11, G12}, {E12, G12} itemsets making 3-itemsets

**3-itemsets**

{C11, D12, E11}=1, {C11, D12, G11}=1, {C11, D12, G12}=1,{C11, E12, G12}=1, { D12, E11, G11}=2, { D12, E11, G12}=0, {D12, E12, G11}=0,{D12, E12, G12}=1,
Finally we find 3-itemset {D12, E11, G11}.

4.2 Comparison

Comparison between the frequent itemsets

Table-A

| Candidate frequent itemset at Level-1 | Algorithm ML_T2L1 | Algorithm in this paper |
|---|---|---|
| 1- itemsets | {1**} {2**} | {{B**}, {C**}, {D**}, {E**}, {F**}, {G**}, {H**} |
| 2-itemsets | {1**, 2**} | {B**, C**}, {C**, D**}, {C**, E**}, {C**, G**}, {D**, E**}, {D**, G**}, {E**, F**}, {E**, G**}, {E**, H**} |
| 3-itemsets | nil | {C**, D**, E**}, {C**, D**, G**}, {D**, E**, G**} |
| 4-itemsets | nil | {C**, D**, E**, G**} |

Table-B

| Candidate frequent itemset at Level-2 | Algorithm ML_T2L1 | Algorithm in this paper |
|---|---|---|
| 1- itemsets | {11*},{12*} {21*}, {22*} | {C1*}, {D1*}, {E1*}, {F1*}, {G1*} |
| 2-itemsets | {11*,12*} {11*, 21*} {11*, 22*} {12*, 22*} {21*, 22*} | {C1*, D1*}, {C1*, E1*}, {C1*, G1*}, {D1*, E1*}, {D1*, G1*}, {E1*, F1*}, {E1*, G1*}, {F1*, G1*} |
| 3-itemsets | {11*,12*, 22*} | {C1*, D1*, E1*}, {C1*, D1*, G1*}, {D1*, E1*, G1*} |
| 4-itemsets | nil | {C1*, D1*, E1*, G1*} |

Table-c

| Candidate frequent itemset at Level-3 | Algorithm ML_T2L1 | Algorithm in this paper |
|---|---|---|
| 1- itemsets | {111},{211},{221} | {C11}, {C12}, {D11}, {D12}, {E11}, {E12}, {G11}, {G12} |
| 2-itemsets | {111, 211} | {C11, D12}, {D12, E11}, {D12, G11}, {E11, G12}, {E12, G12} |
| 3-itemsets | nil | {D12, E11, G11} |

The number of the candidate frequent itemsets at lower level with low mininmum suport and times of scanning the database are compared between this algorithm and ML_T2L1, which is the key player of the efficiency in an algorithm. The algorithm in this paper is better than ML_T2L1 in reducing the time for scanning the database and reduces CPU overhead, reduce memory size. Because we are using princer search method in our algorithm, so by princer search method we reduce the database size and maximum frequent itemset will be generated and written to disk in order to leave space for processing the next frequent itemsets and so on. The database will be scanned k times find out frequent k-itemsets in ML_T2L1 while only two times in our algorithm by putting forward a concept of candidate frequent itemset.

In Table A, B, C the frequent 1-itemsets, 2-itemsets and 3-itemsets for the two algorithms are listed,

## Conclusions

An efficient way to discover the maximum frequent set which can be very useful in various data mining problems, such as the discovery of the multilevel association rules, strong rules and candidate generation, large number of database scans. The maximum frequent set provides an exclusive representation of all the frequent itemsets. In many situations, it suffices to discover the maximum frequent set, and once it is known, all the required frequent subsets can be easily generated. In this paper, we presented an efficient algorithm that can easily discover the maximum frequent set. Our multilevel Pincer-Search algorithm could reduce both the number of times the database is read and the number of candidates considered and reduces the memory requirements significantly and thus makes it possible to process large amounts of data in main memory which in turn reduces I/O operations and makes processing faster.

## Acknowledgments

This work was support and guidance of many people. It is my pleasure to take the opportunity to thank all those who helped me directly or indirectly in preparation of this paper.
I truthfully believe that it is through the godly fruits of my karma that I was blessed with Dr. K.R. Pardasani as my PhD supervisor. I am deeply grateful to him for this tremendous guidance, invaluable suggestions and constant support.

## References

[1] Agrawal R., R. Srikant, "Fast algorithms for mining association rules," In Proceedings of the 20th VLDB Conference, 1999, pp. 487-499.
[2] Agarwal R., Agarwal C. and Prasad V, "A tree projection algorithm for generation of frequent itemsets", Journal of Parallel and Distributed Computing, 2001.
[3] Agarwal R., Srikant R., "Fast Algorithms for Mining Association Rules," Int. Conf. Very Large Data Bases, Santiago, Chile, 1994, pp. 487-499.
[4] Agrawal R., Imielinski T. and Swami A., "Mining association rules between sets of items in large databases," In Proceeding ACM SIGMOD Conference, pp. 207-216, 1993.
[5] Agrawal Rakesh, Srikant Ramakrishnan, "Mining Quantitative Association Rules in Large Relational Tables" Department of Computer Science, University of Wisconsin, Madison, 2004, pp: 809-827.
[6] Bo Yin, Wan, Liang Yong, Ding Li-Ya, "Mining Multilevel Association Rules with Dynamic concept Hierarchy", Proceedings of the Seventh International Conference on Machine Learning and Cybernetics, Kunming, 2008, pp.287-292.

[7] Chen An, Hui Lin Ye. "multiple-level sequential pattern discovery from customer transaction databases," international journal computational intelligence, 2003, pp.48-56.
[8] D. Lin and Z. Kedem. Pincer-Search: A new algorithm for discovering the maximum frequent set. In Proc. 6$^{th}$ EDBT, 1998
[9] Fortin Scott, Liu Ling., "An object-oriented approach to multi-level association rule mining," Proceedings of the fifth international conference on Information and knowledge management,1996, pp.65-72.
[10] Han Jaiwei, Yongjian,"Discovery of multiple level association rule from large database, VLDB Conference, 1995 pp. 420-423.
[11]. Han Jaiwei, Kamber Micheline, "Data Mining Concepts and Techniques," book, 2006.
[12] H. Mannila and H. Toivonen. Discovering frequent episodes in sequences. In Proc. 1st KDD, Aug. 1995.
 [13]  Liu Hunbing and Wang Baishen, "An association Rule Mining Algorithm Based On a Boolean Matrix,"  Data Science Journal, 2007,pp: 559-563.
[14] Stanisic Predrag, Tomovic Savo, "Apriori multiple algorithms for mining association rules," Information technology and control, 2008, pp. 311-320.
[15] Srikant Ramakrishnan, Vu Quoc and Agrawal Rakesh, "Mining Association rules with Item Constraints," Published in KDD Proceedings, 1997.
[16] Pujari Arun K, "Data mining Techniques," Universities press, (India), Hyderabad, 2001.
[17] Pratima Gautam Neelu Khare and K. R. Pardasani, "A model for mining multilevel fuzzy association rule in database," Journal of computing, 2010, pp. 58-68.
[18] Scott Fortin,Ling Liu, "An object-oriented approach to multi-level association rule mining," Proceedings of the fifth international conference on Information and knowledge management, 1996, pp.65-72.
[19] R.S Thakur, R.C. Jain, K.R.Pardasani, "Fast Algorithm for
Mining Multilevel Association Rule Mining," Journal of Computer Science, 2007, pp. 76-81.
[20]Yin –Bo Wan, Yong Liang and Li-Ya Ding, "Mining Multilevel Association rules with dynamic concept hierarchy," IEEE,  2008, pp. 287-292.

Dr. Pratima Gautam received the Ph.D. degree from Maulana Azad National Institute of Technology (Manit) deemed university, Bhopal in 2012; she is working as a Professor in the Department of computer science and engineering Manit (Bhopal), India. She has 12 publications in National and International Journals.